\documentclass[10pt,twocolumn,pre,byrevtex,showpacs]{revtex4}
\usepackage{amssymb}
\usepackage{amsmath}
\usepackage{epsfig}

\begin{document}

\title{Information-theoretic approach to the study of control systems}

\author{Hugo Touchette}

\affiliation{\mbox{Department of Physics and School of Computer Science,
McGill University, Montr\'eal, Qu\'ebec, Canada H3A 2A7}}

\author{Seth Lloyd}

\affiliation{d'Arbeloff Laboratory for Information Systems and Technology, Department
                  of Mechanical Engineering, Massachusetts Institute of Technology,
                  Cambridge, Massachusetts 02139}

\date{\today}

\begin{abstract}
We propose an information-theoretic framework for analyzing control systems
based on the close relationship of controllers to communication channels. A
communication channel takes an input state and transforms it into an output
state. A controller, similarly, takes the initial state of a system to be
controlled and transforms it into a target state. In this sense, a
controller can be thought of as an actuation channel that acts on inputs to
produce desired outputs. In this transformation process, two different
control strategies can be adopted: (i) the controller applies an actuation
dynamics that is independent of the state of the system to be controlled
(open-loop control); or (ii) the controller enacts an actuation dynamics
that is based on some information about the state of the controlled system
(closed-loop control). Using this communication channel model of control, we
provide necessary and sufficient conditions for a system to be perfectly
controllable and perfectly observable in terms of information and entropy.
In addition, we derive a quantitative trade-off between the amount of
information gathered by a closed-loop controller and its relative
performance advantage over an open-loop controller in stabilizing a system.
This work supplements earlier results [H. Touchette, S. Lloyd,
Phys.~Rev.~Lett.~\textbf{84}, 1156 (2000)] by providing new derivations of
the advantage afforded by closed-loop control and by proposing an
information-based optimality criterion for control systems. New applications
of this approach pertaining to proportional controllers, and the control of
chaotic maps are also presented.
\end{abstract}

\pacs{02.50.Ey, 05.90.+m, 89.70.+c, 07.05.Dz, 05.45.Gg}

\maketitle

\section{Introduction}

It is common in studying controllers to describe the interplay between the 
\textit{sensors} which estimate the state of a system intended to be
controlled, and the \textit{actuators} used to actually modify the dynamics
of the controlled system as a transfer of information involving three steps:
estimation, decision, and actuation. In the first step, sensors are used to
gather information from the controlled system in the form of data relative
to its state (estimation step). This information is then processed according
to some plan or control strategy in order to determine which control
dynamics is to be applied (decision step), to be finally transferred to the
actuators which feed the processed information back to the controlled system
to modify its dynamics, typically with the goal of decreasing the
uncertainty in the value of the system's variables (actuation step) \cite
{wiener1948,dazzo1988,singh1987}. 

Whether or not the estimation step is
present in this sequence is optional, and determines which type of control
strategy is used. In so-called \textit{closed-loop} or \textit{feedback}
control techniques, actuators rely explicitly on the information provided by
sensors to apply the actuation dynamics, whereas in \textit{open-loop}
control there is no estimation step preceding the actuation step. In other
words, an open-loop controller distinguishes itself from a closed-loop
controller in that it does not need a continual input of `selective'
information \cite{mackay1969} to work: like a throttle or a hand brake, it
implements a control action independently of the state of the controlled
system. In this respect, open-loop control techniques represent a subclass
of closed-loop controls that neglect the information made available by
estimation.

Since control is fundamentally about information (getting it, processing it,
and applying it) it is perhaps surprising to note that few efforts have been
made to develop a quantitative theory of controllers focused on a clear and
rigorous definition of information. Indeed, although controllers have been
described by numerous authors as information gathering and using systems
(see, e.g., \cite{wiener1948,scano1965,ashby1956,ashby1965}), and despite
many results related to this problem \cite
{sankoff1965,weidemann1969,pop1975,pop1980,wein1982,saridis1988,saridis21988,saridis1995,saridis1997,del1989,pande1990,lloyd1996,mitter1997,sahai1999,mitter2000}
, there exists at present no general information-theoretic formalism
characterizing the exchange of information between a controlled system and a
controller, and more importantly, which allows for the assignation of a
definite value of information in control processes \cite
{belis1968,antsaklis2000}. To address this deficiency, we present in this
paper with a quantitative study of the role of information in control. The
basis of the results presented here was first elaborated first in \cite
{hugo12000}, and draws upon the work of several of the papers cited above by
bringing together some aspects of dynamical systems, information theory, in
addition to probabilistic networks to construct control models in the
context of which quantities analogous to entropy can be defined.

Central to our approach is the notion of a communication channel, and its
extension to the idea of \textit{control channels}. As originally proposed
by Shannon \cite{shannon1948}, a (memoryless) communication channel can be
represented mathematically by a probability transition matrix, say $p(y|x)$,
relating the two random variables $X$ and $Y$ which are interpreted,
respectively, as the input and the output of the channel. In the next two
sections of the present work, we adapt this common probabilistic picture of
communication engineering to describe the operation of a basic control
setup, composed of a sensor linked to an actuator, in terms of two channels:
one coupling the initial state of the system to be controlled and the state
of the sensor (sensor channel), and another one describing the state
evolution of the controlled system as influenced by the sensor-actuator's
states (actuation channel).

In Sections IV and V, we use this model in conjunction with the properties
of entropy-like quantities to exhibit fundamental results pertaining to
control systems. As a first of these results, we show that the classical
definition of controllability, a concept well-known to the field of control
theory, can be rephrased in an information-theoretic fashion. This
definition is used, in turn, to show that a system is perfectly controllable
upon the application of controls if, and only if, the target state of that
system is statistically independent of any other external systems playing
the role of noise sources. A similar information-theoretic result is also
derived for the complementary concept of observability. Moreover, we provide
bounds on the amount of information a feedback controller must gather in
order to stabilize the state of a system. More precisely, we prove that the
amount of information gathered by the controller must be bounded below by
the difference $\Delta H_{\text{closed}}-\Delta H_{\text{open}}^{\max }$,
where $\Delta H_{\text{closed }}$ is the closed-loop entropy reduction that
results from utilizing information in the control process, and $\Delta H_{
\text{open}}^{\max }$ is the maximum decrease of entropy attainable when
restricted to open-loop control techniques. This last result, as we will
see, can be used to define an information-based optimality criterion for
control systems.

The idea of reducing the entropy of a system using information gathered from
estimating its state is not novel by itself. Indeed, as he wondered about
the validity of the second law of thermodynamics, the physicist James Clerk
Maxwell was probably the first to imagine in 1897 a device (or a `demon' as
it was later called) whose task is to reduce the entropy of a gas using
information about the positions and velocities of the particles forming the
gas. (See \cite{leff1990} for a description of Maxwell's demon and a guide
to this subject's literature.) In the more specific context of control
theory, the problem of reducing the entropy of a dynamical system has also
been investigated, notably by Poplavski\u{\i} \cite{pop1975,pop1980} and by
Weidemann \cite{weidemann1969}. Poplavski\u{\i} analyzed the information
gathered by sensors in terms of Brillouin's notion of negentropy \cite
{leff1990,brillouin1956}, and derived a series of physical limits to
control. His study focuses on the sensor part of controllers, leaving aside
the actuation process which, as will be shown, can be also treated in an
information-theoretic fashion. In a similar way, Weidemann performed an
information-based analysis of a class of linear controllers having measure
preserving sensors. Other related ideas and results can be found in Refs.
\cite
{sankoff1965,wein1982,saridis1988,saridis21988,saridis1995,saridis1997,del1989,pande1990,lloyd1996,mitter1997,sahai1999,mitter2000}
.

In the present paper, we build on these studies and go further by presenting
results which apply equally to linear and nonlinear systems, and can be
generalized with the aid of a few modifications to encompass
continuous-space systems as well as continuous-time dynamics. To illustrate
this scope of applications, we study in Section VI specific examples of
control systems. Among these, we consider two variants of proportional
controllers, which play a predominant role in the design of present-day
controllers, in addition to complete our numerical investigation of
noise-perturbed chaotic controllers initiated in \cite{hugo12000}. Finally,
we remark in Section VII on the relationship of our framework with
thermodynamics and optimal control theory.

\section{Channel-like models of control}

\begin{figure}[t]

\epsfig{file=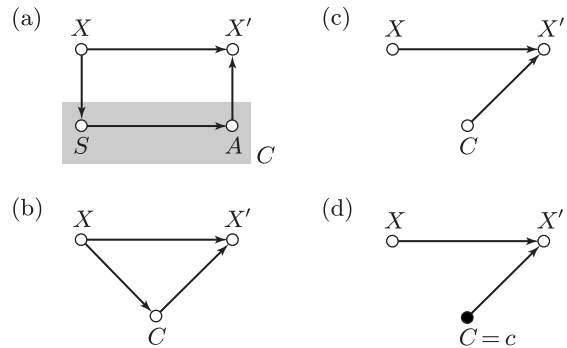,clip=}

\caption{Directed acyclic graphs representing a basic control
process. (a) Full control system with a sensor $S$ and an actuator $A$.
(b) Reduced closed-loop diagram obtained by merging the sensor and the actuator
into a single controller device, the controller. (c) Reduced open-loop control diagram.
(d) Single actuation channel enacted by the controller's state $C=c$.}

\end{figure}

In this section, we introduce a simple control model that allows
investigation of the dynamical interplay that exists between a sensor and an
actuator to `move' a system from an unknown initial state to a desired final
target state. Such a process is depicted schematically in Figure 1 in the
form of directed acyclic graphs, also known as Bayesian networks \cite
{pearl1988,jordan1999}. The vertices of these graphs correspond to random
variables representing the state of a (classical) system; the arrows give
the probabilistic dependencies among the random variables according to the
general decomposition 
\begin{equation}
p(x_1,x_2,\ldots ,x_N)=\prod_{i=1}^Np(x_i|\pi [X_i]),  \label{dec1}
\end{equation}
where $\pi [X_i]$ is the set of random variables which are direct parents of 
$X_i$, $i=1,2,\ldots ,N$, ($\pi [X_1]=\emptyset $). The acyclic condition of
the graphs ensures that no vertex is a descendant or an ancestor of itself,
in which case we can order the vertices chronologically, i.e., from
ancestors to descendants. This defines a causal ordering, and, consequently,
a time line directed on the graphs from left to right.

In the control graph of Figure 1a, the random variable $X$ represents the
initial state of the system to be controlled, and whose values $x\in 
\mathcal{X}$ are drawn according to a fixed probability distribution $p_X(x)$
. In conformity with our introductory description of controllers, this
initial state is controlled to a final state $X^{\prime }$ with state values 
$x^{\prime }\in \mathcal{X}$ by means of a sensor, of state variable $S$,
and an actuator whose state variable $A$ influences the transition from $X$
to $X^{\prime }$. For simplicity, all the random variables describing the
different systems are taken to be discrete random variables with finite sets
of outcomes. The extension to continuous-state systems is discussed in
Section IV. Also, to further simplify the analysis of this model, we assume
throughout this paper that the sensor and the actuator are merged into a
single device, called the \textit{controller}, which fulfills both the roles
of estimation and actuation (see Figure 1b). The state of the controller is
denoted by $C$, and assumes values from some set $\mathcal{C}$ of admissible
controls \cite{note1}.

Using this notation together with the decomposition of Eq.(\ref{dec1}), the
joint distribution $p(x,x^{\prime },c)$ describing the causal dependencies
between the states of the control graphs can now be constructed. For
instance, the complete joint distribution corresponding to the closed-loop
graph of Figure 1b is written as 
\begin{equation}
p(x,x^{\prime },c)_{\text{closed}}=p_X(x)p(c|x)p(x^{\prime }|x,c),
\label{cld}
\end{equation}
while the open-loop version of this graph, depicted in Figure 1c, is
characterized by a joint distribution of the form 
\begin{equation}
p(x,x^{\prime },c)_{\text{open}}=p_X(x)p_C(c)p(x^{\prime }|x,c).  \label{opd}
\end{equation}
Following the definition of closed- and open-loop control given above, what
distinguishes probabilistically and graphically both control strategies is
the presence, for closed-loop control, of a direct correlation link between $
X$ and $C$ represented by the conditional probability $p(c|x)$. This
correlation can be thought of as a (possibly noisy) communication channel,
referred here to as the \textit{sensor} or \textit{measurement} channel,
that enables the controller to gather an amount of information identified
formally with the \textit{mutual information} 
\begin{equation}
I(X;C)=\sum_{x\in \mathcal{X},c\in \mathcal{C}}p_{XC}(x,c)\log \frac{
p_{XC}(x,c)}{p_X(x)p_C(c)},  \label{mutual}
\end{equation}
where $p_{X,C}(x,c)=p_X(x)p(c|x)$. (All logarithms are assumed to the base
2, except where explicitly noted.) Recall that $I(X;C)\geq 0$ with equality
if and only if the random variables $X$ and $C$ are statistically
independent \cite{cover1991}, so that in view of this quantity we are
naturally led to define open-loop control with the requirement that $
I(X;C)=0 $; closed-loop control, on the other hand, must be such that $
I(X;C)\neq 0$.

As for the actuation part of the control process, the joint distributions of
Eqs.(\ref{cld})-(\ref{opd}) show that it is accounted for by the
channel-like probability transition matrix $p(x^{\prime }|x,c)$. The entries
of this \textit{actuation} matrix give the probability that the controlled
system in state $X=x$ is actuated to $X^{\prime }=x^{\prime }$ given that
the controller's state is $C=c$. From here on, it will be convenient to
think of the control actions indexed by each value of $C$ as a set of 
\textit{\ actuation channels}, with memoryless transition matrices 
\begin{equation}
p(x^{\prime }|x)_c=p(x^{\prime }|x,c),
\end{equation}
governing the transmission of the random variable $X$ to a target state $
X^{\prime }$. In terms of the control graphs, such channels are represented
in the same form as in Figure 1d to show that the fixed value $C=c$ (filled
circle in the graph) enacts a transformation of the random variable $X$
(open circle) to a yet unspecified value associated with the random variable 
$X^{\prime }$ (open circle as well). Guided by this graphical
representation, we will show in the next section that the overall action of
a controller can be decomposed into a series of single conditional actuation
actions or \textit{subdynamics} triggered by the internal state of $C$.

Here we characterize the effect of the subdynamics available to a controller
on the \textit{entropy} of the initial state $X$: 
\begin{equation}
H(X)=-\sum_{x\in \mathcal{X}}p_X(x)\log p_X(x).
\end{equation}
In theory, this effect is completely determined by the choice of the initial
state $X$, and the form of the actuation matrices. The effect of these two
`variables' on $H(X)$ is categorized according to the three following
classes of dynamics:

\textit{One-to-one transitions}: A given control subdynamics specified by $
C=c$ conserves the entropy of the initial state $X$ if the corresponding
probability matrix $p(x^{\prime }|x)_c$ is that of a noiseless channel.
Permutations or translations of $X$ are examples of this sort of dynamics.

\textit{Many-to-one transitions}:~A control channel $p(x^{\prime }|x)_c$ may
cause some subset $\mathcal{X}_c$ of the state space $\mathcal{X}$ to be
mapped onto a smaller subset of values for $X^{\prime }$. In this case, the
corresponding subdynamics is said to be \textit{dissipative} or \textit{\
volume-contracting} as it decreases the entropy of ensembles of states lying
in $\mathcal{X}_c$.

\textit{One-to-many transitions}:~A channel $p(x^{\prime }|x)_c$ can also
lead $H(X)$ to increase if it is \textit{non-deterministic}, i.e., if it
specifies the image of one or more values of $X$ only up to a certain
probability different than zero or one. This will be the case, for example,
if the actuator is unable to accurately manipulate the dynamics of the
controlled system, or if any part of the control system is affected by
external and non-controllable systems.

From a strict mathematical point of view, note that any non-deterministic
channel modeling a source of noise at the level of actuation or estimation
can be represented abstractly as a randomly selected deterministic channel
with transition matrix containing only zeros and ones. The outcome of a
random variable undisclosed to the controller can be thought of as being
responsible for the choice of the channel to use. Figure 2 shows
specifically how this can be done by supplementing our original control
graphs of Figure 1 with an exogenous and non-controllable random variable $Z$
in order to `purify' the channel considered (actuation or estimation) \cite
{note2}. For the actuation channel, as for instance, the purification
condition simply refers to the two following properties:

(i) The mapping from $X$ to $X^{\prime }$ conditioned on the values $c$ and $
z$, as described by the extended transition matrix $p(x^{\prime }|x,c,z)$,
is deterministic for all $c\in \mathcal{C}$ and $z\in \mathcal{Z}$;

(ii) When traced out of $Z$, $p(x^{\prime }|x,c,z)$ reproduces the dynamics
of $p(x^{\prime }|x,c)$, i.e., 
\begin{equation}
p(x^{\prime }|x,c)=\sum_{z\in \mathcal{Z}}p(x^{\prime }|x,c,z)p_Z(z),
\end{equation}
for all $x^{\prime }$, $x\in \mathcal{X}$, and all $c\in \mathcal{C}$.

\begin{figure}[t]

\epsfig{file=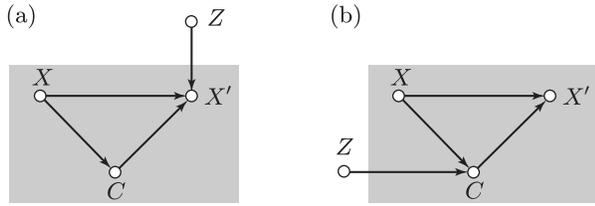,clip=}

\caption{Control diagrams illustrating the purification procedure for (a) the actuation
channel, and (b) the sensor channel. Purifying a channel, for instance the sensor
channel, simply means that knowing the value of $X$ and $Z$ enables one to know 
with probability one the value of $C$. However, discarding (viz, tracing out) any 
information concerning $Z$ leaves us with some uncertainty as to which $C$ is
reached from a given value for $X$.}

\end{figure}

\section{Conditional analysis}

To complement the material introduced in the previous section, we now
present a technique for analyzing the control graphs that emphasizes further
the conceptual importance of the actuation channel and its graphical
representation. The technique is based on a useful symmetry of Figure 1c
that enables us to separate the effect of the random variable $X$ in the
actuation matrix from the effect of the control variable $C$. From one
perspective, the open-loop decomposition 
\begin{equation}
p_{X^{\prime }}(x^{\prime })_{\text{open}}=\sum_cp_C(c)\left[
\sum_xp(x^{\prime }|x,c)p_X(x)\right]  \label{opds}
\end{equation}
suggests that an open-loop control process can be decomposed into an
ensemble of actuations, each one indexed by a particular value $c$ that
takes the initial distribution $p_X(x)$ to a conditional distribution (first
sum in parentheses) 
\begin{equation}
p(x^{\prime }|c)_{\text{open}}=\sum_{x\in \mathcal{X}}p(x^{\prime
}|x,c)p_X(x).  \label{copd}
\end{equation}
The final marginal distribution $p_{X^{\prime }}(x^{\prime })_{\text{open}}$
is then obtained by evaluating the second sum in Eq.(\ref{opds}), thus
averaging $p(x^{\prime }|c)_{\text{open}}$ over the control variable. From
another perspective, Eq.(\ref{opds}), re-ordered as 
\begin{equation}
p_{X^{\prime }}(x^{\prime })_{\text{open}}=\sum_xp_X(x)\left[
\sum_cp(x^{\prime }|x,c)p_C(c)\right] ,
\end{equation}
indicates that the overall action of a controller can be seen as
transmitting $X$ through an `averaged' channel (sum in parentheses) whose
transition matrix is given by 
\begin{equation}
p(x^{\prime }|x)=\sum_{c\in \mathcal{C}}p(x^{\prime }|x,c)p_C(c).
\end{equation}
In the former perspective, each actuation subdynamics represented by the
control graph of Figure 1d can be characterized by a \textit{conditional
open-loop entropy reduction} defined by 
\begin{equation}
\Delta H_{\text{open}}^c=H(X)-H(X^{\prime }|c)_{\text{open}}  \label{opc}
\end{equation}
where 
\begin{equation}
H(X^{\prime }|c)=-\sum_{x^{\prime }\in \mathcal{X}}p(x^{\prime }|c)\log
p(x^{\prime }|c).  \label{cent1}
\end{equation}
(Subscripts of $H$ indicate from which distribution the entropy is to be
calculated.) In the latter perspective, the entropy reduction associated
with the unconditional transition from $X$ to $X^{\prime }$ is simply the 
\textit{open-loop entropy reduction} 
\begin{equation}
\Delta H_{\text{open}}=H(X)-H(X^{\prime })_{\text{open}}  \label{oper}
\end{equation}
which characterizes the control process as a whole, without regard to any
knowledge of the controller's state.

For closed-loop control, the decomposition of the control action into a set
of conditional actuations seems \textit{a priori} inapplicable, for the
controller's state itself depends on the initial state of the controlled
system, and thus cannot be fixed at will. Despite this fact, one can use the
Bayesian rule of statistical inference 
\begin{equation}
p(x|c)=\frac{p(c|x)p_X(x)}{p_C(c)},
\end{equation}
where 
\begin{equation}
p_C(c)=\sum_{x\in \mathcal{X}}p(c|x)p_X(x),  \label{tpc1}
\end{equation}
to invert the dependency between $X$ and $C$ in the sensor channel so as to
rewrite the closed-loop decomposition in the following form: 
\begin{equation}
p_{X^{\prime }}(x^{\prime })_{\text{closed}}=\sum_cp_C(c)\left[
\sum_xp(x^{\prime }|x,c)p(x|c)\right] .  \label{clpds}
\end{equation}
By comparing this last equation with Eq.(\ref{opds}), we see that a
closed-loop controller is essentially an open-loop controller acting on the
basis of $p(x|c)$ instead of $p_X(x)$ \cite{ho1964}. Thus, given that $c$ is
fixed, a closed-loop equivalent of Eq.(\ref{opc}) can be calculated simply
by substituting $p_X(x)$ with $p(x|c)$, thereby obtaining 
\begin{equation}
\Delta H_{\text{closed}}^c=H(X|c)-H(X^{\prime }|c)  \label{clerc}
\end{equation}
for all $c$.

\begin{figure*}[t]

\epsfig{file=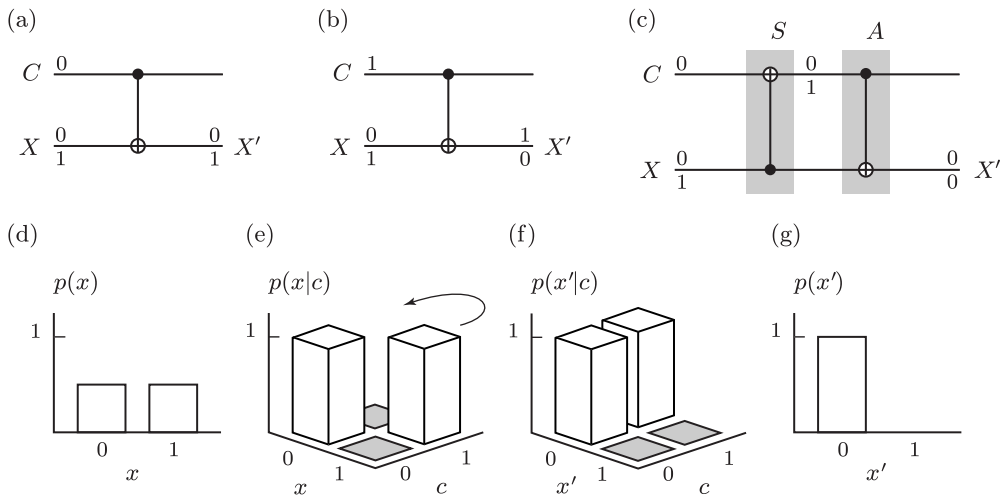,clip=}

\caption{Controlled-{\sc not} controller. (a) Boolean circuit illustrating
the effect of the controller's state $C=0$ on the input states ${0,1}$ of
the controlled system $X$ (identity in this case). (b) Control action 
triggered by $C=1$ (swapping).
(c) Complete control system with sensor $S$ and actuator $A$. Note that
the sensor itself is modeled by a {\sc cnot} gate. (d)-(g) State of the controlled
system at  different stages of the control depicted in the spirit of conditional analysis. 
(d) A uniformly distributed input state $X$ is measured by the sensor in 
such a way that the conditional random variable $X|c$ is deterministic (e). 
(f) The control action triggered by $C$ has the effect of swapping the 
values $x$ for which $C=1$. (g) Deterministic probability distribution for the final state 
$X'$ upon averaging over $C$.}

\end{figure*}

The rationale for decomposing a closed-loop control action into a set of
conditional actuations can be justified by observing that \textit{a
closed-loop controller, after the estimation step, can be thought of as an
ensemble of open-loop controllers acting on a set of estimated states}. In
other words, what differentiates open-loop and closed-loop control from the
viewpoint of the actuator is the fact that, for the former strategy, a given
control action selected by $C=c$ transforms all the values $x$ contained in
the \textit{support} of $X$, i.e., the set 
\begin{equation}
\text{supp}(X)=\{x\in \mathcal{X}:p_X(x)>0\},
\end{equation}
whereas for the latter strategy, namely closed-loop control, the same
actuation only affects the support of the posterior distribution $p(x|c)$
associated with $X|c$, the random variable $X$ conditioned on the outcome $c$
. This is so because the decision as to which control value is used has been
determined according to the observation of specific values of $X$ which are
in turn affected by the chosen control value. By combining the influence of
all the control values, we thus have that information gathered by the sensor
affects the entire control process by inducing a \textit{covering} of the
support space 
\begin{equation}
\text{supp}(X)=\bigcup_{c\in \mathcal{C}}\text{supp}(X|c),  \label{cov1}
\end{equation}
in such a way that values $x\in \text{supp}(X|c_1)$, for a fixed $c_1\in 
\mathcal{C}$, are controlled by the corresponding actuation channel $
p(x^{\prime }|x,C=c_1)$, while other values in $\text{supp}(X|c_2)$ are
controlled using $p(x^{\prime }|x,C=c_2)$, and so on for all $c_i\in 
\mathcal{C}$. This is manifest if one compares Eqs.(\ref{opds}) and (\ref
{clpds}). Note that a particular value $x$ included in $\text{supp}(X)$ may
be actuated by many different control values if it is part of more than one
`conditional' support $\text{supp}(X|c)$. Hence the fact that Eq.(\ref{cov1}
) only specifies a covering, and not necessarily a partition constructed
from non-overlapping sets. Whenever this occurs, we say that the control is 
\textit{mixing}.

To illustrate the above ideas about subdynamics applied to conditional
subsets of $\mathcal{X}$ in a more concrete setting, we proceed in the next
paragraph with a basic example involving the control of a binary state
system using a controller restricted to use permutations as actuation rules 
\cite{hugo12000}. This example will be used throughout the article as a test
situation for other concepts.

\textit{Example 1.} Let $C$ be a binary state controller acting on a bit $X$
by means of a so-called controlled-\textsc{not} (\textsc{cnot}) logical
gate. As shown in the circuits of Figures 3a-b, the state $X$, under the
action of the gate, is left intact or is negated depending on the control
value: 
\begin{equation}
x^{\prime }=\left\{ 
\begin{array}{ll}
x, & \text{if }c=0 \\ 
x\oplus 1, & \text{if }c=1.
\end{array}
\right.
\end{equation}
($\oplus $ stands for modulo $2$ addition.) Furthermore, assume that the
controller's state is determined by the outcome of a `perfect' sensor which
can be modeled by another \textsc{cnot} gate such that $C=X$ when $C$ is
initially set to $0$ (Figure 3c). As a result of these actuation rules, it
can be verified that $\Delta H_{\text{open}}^c=\Delta H_{\text{closed}}^c=0$
, and so the application of a single open- or closed-loop control action
cannot increase the uncertainty $H(X)$. In fact, whether the subdynamics is
applied in an open- or closed-loop fashion is irrelevant here: a permutation
is just a permutation in either cases. Now, since $C=X$, we have that the
random variable $X$ conditioned on $C=c$ must be equal to $c$ with
probability one. For closed-loop control, this implies that the value $X=0$,
which is the only element of $\text{supp}(X|C=0)$, is kept constant during
actuation, whereas the value $X=1$ in $\text{supp}(X|C=1)$ is negated to $0$
in accordance with the controller's state $C=1$ (Figure 3e). Under this
control action, the conditional random variable $X^{\prime }|c$ is forced to
assume the same deterministic value for all $c$, implying that $X^{\prime }$
must be deterministic as well, regardless of the statistics of $C$ (Figures
3f-g). Therefore, $H(X^{\prime })_{\text{closed}}=0$. In contrast, the
application of the same actuation rules in an open-loop fashion transform
the state $X$ to a final state having, at best, no less uncertainty than
what is initially specified by the statistics of $X$, i.e., $H(X^{\prime })_{
\text{open}}\geq H(X)$.\hfill $\blacksquare $

\section{Entropic formulation of controllability and observability}

The first instance of the general control problem that we now proceed to
study involves the dual concepts of controllability and observability. In
control theory, the importance of these concepts arises from the fact that
they characterize mathematically the input-output structure of a system
intended to be controlled, and thereby determine whether a given control
task is realizable or not \cite{dazzo1988,singh1987}. In short,
controllability is concerned with the possibilities and limitations of the
actuation channel or, in other words, the class of control dynamics that can
be effected by a controller. Observability, on the other hand, is concerned
with the set of states which are accessible to estimation given that a
particular sensor channel is used. In this section, prompted by preliminary
results obtained by Lloyd and Slotine \cite{lloyd1996}, we define entropic
analogs of the widely held control-theoretic definitions of controllability
and observability, and explore the consequences of these new definitions.

\subsection{Controllability}

In its simplest expression, a system is said to be \textit{controllable} at $
X=x$ if any of the final state $X^{\prime }=x^{\prime }$ can be reached from 
$X=x$ using at least one control input $C=c$ \cite{dazzo1988,singh1987}.
Allowing for non-deterministic control actions, we may refine this
definition and say that a system is \textit{perfectly controllable} at $X=x$
if it is controllable at $X=x$ with probability 1, i.e., if, for any $
x^{\prime }$, there exists at least one $c$ such that $p(x^{\prime }|x,c)=1$
. In other words, a system is perfectly controllable if (i) all final states
for $X^{\prime }$ are reachable from $X=x$ (complete reachability
condition); and (ii) all final states for $X^{\prime }$ are connected to $
X=x $ by at least one deterministic subdynamics (deterministic transitions
condition). In terms of entropy, these two conditions are translated as
follows. (The next result was originally put forward in \cite{lloyd1996}
without a complete proof.)

\textit{Theorem 1.} A system is perfectly controllable at $X=x$ if and only
if there exists a distribution $p(c|x)$ \cite{note3} such that 
\begin{equation}
p(x^{\prime }|x)=\sum_{c\in \mathcal{C}}p(x^{\prime }|x,c)p(c|x)\neq 0
\end{equation}
for all $x^{\prime }$, and 
\begin{equation}
H(X^{\prime }|x,C)=\sum_{c\in \mathcal{C}}H(X^{\prime }|x,c)p(c|x)=0,
\label{cec}
\end{equation}
where 
\begin{equation}
H(X^{\prime }|x,c)=-\sum_{x^{\prime }\in \mathcal{X}}p(x^{\prime }|x,c)\log
p(x^{\prime }|x,c).
\end{equation}

\textit{Proof.} If $x$ is controllable, then for each $x^{\prime }$ there
exists at least one control value $c=c(x^{\prime },x)\in \mathcal{C}$ such
that $p(x^{\prime }|x,c)=1$, and thus $H(X^{\prime }|x,c)=0$. Also, choosing 
\begin{equation}
\text{supp}(C|x)=\{c:p(x^{\prime }|x,c)=1\}
\end{equation}
over all $x^{\prime }\in \mathcal{X}$ and $X=x$ ensures that the average
conditional entropy over the conditional random variable $C|x$ vanishes, and
that $p(x^{\prime }|x)\neq 0$. This proves the direct part of the theorem.
To prove the converse, note that if $p(x^{\prime }|x)\neq 0$ for a given $
x^{\prime }$, then there is at least one value $c$ for which $p(x^{\prime
}|x,c)\neq 0$, which means that there is at least one subdynamics connecting 
$x$ to $x^{\prime }$. If in addition we have $H(X^{\prime }|x,C)=0$, then we
can conclude that such a subdynamics must in fact be deterministic. As this
is verified for any state value $x^{\prime }$, we obtain in conclusion that
for all $x^{\prime }\in \mathcal{X}$ there exists a $c$ such that $
p(x^{\prime }|x,c)=1$.\hfill $\blacksquare $

In the case where a system is only \textit{approximately controllable},
i.e., controllable but not in a deterministic fashion, the conditional
entropy $H(X^{\prime }|x,C)$ has the desirable feature of being
interpretable as the residual uncertainty or uncontrolled variation left in
the output $X^{\prime }$ when the controller's state $C$ is chosen with
respect to the initial value $x$ \cite{lloyd1996}. If one regards $C$ as an
input to a communication channel and $X^{\prime }$ as the channel output,
then the degree to which the final state $X^{\prime }$ is controlled by
manipulating the controller's state can be identified with the conditional
mutual information $I(X^{\prime };C|x)$. This latter quantity can be
expressed either using a formula similar to Eq.(\ref{mutual}), or by using
the expression 
\begin{equation}
I(X^{\prime };C|x)=H(X^{\prime }|x)-H(X^{\prime }|x,C),  \label{mutualc}
\end{equation}
which is a conditional version of the chain rule 
\begin{equation}
I(X;Y)=H(X)-H(X|Y),  \label{cr1}
\end{equation}
valid for any random variables $X$ and $Y$.

Note that the two above equations allow for another interpretation of $
H(X^{\prime }|x,C)$. The conditional entropy $H(X|Y)$, entering in (\ref{cr1}
), is often interpreted in communication theory as representing an
information loss (the so-called equivocation of Shannon \cite{shannon1948}),
which results from substracting the maximum noiseless capacity $I(X;X)=H(X)$
of a communication channel with input $X$ and output $Y$ from the actual
capacity of that channel as measured by $I(X;Y)$. In our case, we can apply
the same reasoning to Eq.(\ref{mutualc}), and interpret the quantity $
H(X^{\prime }|x,C)$ as a \textit{control loss} which appears as a negative
contribution in the expression of $I(X^{\prime };C|x)$, the number of bits
of accuracy to which specifying the control variable specifies the output
state of the controlled system. This means that higher is the quantity $
H(X^{\prime }|x,C)$, then higher is the uncertainty or imprecision
associated with the outcome of $X^{\prime }$ upon application of the control
action.

\subsection{Complete and average controllability}

In order to characterize the \textit{complete} controllability of a system,
i.e., its controllability properties over all possible initial states,
define 
\begin{eqnarray}
L_C &=&\min_{\{p(c|x)\}}H(X^{\prime }|X,C)  \nonumber \\
&=&\min_{\{p(c|x)\}}\sum_{x\in \mathcal{X}}p_X(x)\sum_{c\in \mathcal{C}
}H(X^{\prime }|x,c)p(c|x)  \label{clss}
\end{eqnarray}
as the \textit{average control loss}. (The minimization over all conditional
distributions for $C$ is there to ensure that $L_C$ reflects the properties
of the actuation channel, and does not depend on one's choice of control
inputs.) With this definition, we have that a system is perfectly
controllable over the support of $X$ if $L_C=0$ and $p(x^{\prime }|x)\neq 0$
for all $x^{\prime }$. In any other cases, it is approximately controllable
for at least one $x$. The proof of this result follows essentially by noting
that, since discrete entropy is positive definite, the condition $
H(X^{\prime }|X,C)=0$ necessarily implies $H(X^{\prime }|x,C)=0$ for all $
x\in \text{supp}(X)$.

The next two results relate the average control loss with other quantities
of interest. Control graphs containing the purification of the actuation
channel, as depicted in Figure 2, are used throughout the rest of this
section.

\textit{Theorem 2.} Under the assumption that $X^{\prime }$ is a
deterministic random variable conditioned on the values $x$, $c$, and $z$
(purification assumption), we have $L_C\leq H(Z)$ with equality if, and only
if, $H(Z|X^{\prime },X,C)=0$.

\textit{Proof.} Using the general inequality $H(X)\leq H(X,Y)$, and the
chain rule for joint entropies, one may write 
\begin{eqnarray}
H(X^{\prime }|X,C) &\leq &H(X^{\prime },Z|X,C)  \nonumber \\
&=&H(Z|X,C)+H(X^{\prime }|X,C,Z).  \label{clin1}
\end{eqnarray}
However, $H(X^{\prime }|X,C,Z)=0$, since the knowledge of the triplet $
(x,c,z)$ is sufficient to infer the value of $X^{\prime }$ (see the
conditions in Section II). Hence, 
\begin{eqnarray}
H(X^{\prime }|X,C) &\leq &H(Z|X,C)  \nonumber \\
&=&H(Z),
\end{eqnarray}
where the last equality follows from the fact that $Z$ is chosen
independently of $X$ and $C$ as illustrated in the control graph of Figure
2a. Now, from the chain rule 
\begin{equation}
H(X^{\prime },Z|X,C)=H(X^{\prime }|X,C)+H(Z|X^{\prime },X,C),
\end{equation}
it is clear that equality in the first line of expression (\ref{clin1}) is
achieved if and only if $H(Z|X^{\prime },X,C)=0$.\hfill$\blacksquare $

The result of Theorem 2 demonstrates that the uncertainty associated with
the control of the state $X$ is upper bounded by the noise level of the
actuation channel as measured by the entropy of $Z$. This agrees well with
the fact that one goal of controllers is to protect a system against the
effects of its environment so as to ensure that it is minimally affected by
noise. In the limit where the control loss vanishes, the state $X^{\prime }$
of the controlled system should show no variability given that we know the
initial state and the control action, even in the presence of actuation
noise, and should thus be independent of the random variable $Z$. This is
the essence of the next two results which hold for the same conditions as
Theorem 2 (the minimization over the set of conditional probability
distributions $\{p(c|x)\}$ is implied at this point).

\textit{Theorem 3.} $L_C=I(X^{\prime };Z|X,C)$.

\textit{Proof.} From the chain rule of mutual information, we can easily
derive 
\begin{equation}
I(X^{\prime };Z|X,C)=H(X^{\prime }|X,C)-H(X^{\prime }|X,C,Z).  \label{clin2}
\end{equation}
Thus, $I(X^{\prime };Z|X,C)=H(X^{\prime }|X,C)$ if we use again the
deterministic property of the random variable $X^{\prime }|x,c,z$ upon
purification of $p(x^{\prime }|x,c)$.\hfill$\blacksquare $

\textit{Theorem 4.} $L_C=I(X^{\prime };X,C,Z)-I(X^{\prime };X,C)$.

\textit{Proof.} Using the chain rule of mutual information, we write 
\begin{eqnarray}
I(X^{\prime };X,C,Z) &=&H(X^{\prime })-H(X^{\prime }|X,C,Z)  \nonumber \\
&=&H(X^{\prime })-H(X^{\prime }|X,C,Z)  \nonumber \\
&&+H(X^{\prime }|X,C)-H(X^{\prime }|X,C)  \nonumber \\
&=&I(X^{\prime };X,C)+I(X^{\prime };Z|X,C).
\end{eqnarray}
For the last equality, we have used Eq.(\ref{clin2}). Now, by substituting $
L_C=I(X^{\prime };Z|X,C)$ from the previous theorem, we obtain the desired
result.\hfill$\blacksquare $

As a direct corollary of these two results, we have that a system is
completely and perfectly controllable if, and only if, $I(X^{\prime };Z|X,C)$
is equal to zero or equivalently if, and only if, 
\begin{equation}
I(X^{\prime };X,C,Z)=I(X^{\prime };X,C).
\end{equation}
Hence, a necessary and sufficient entropic condition for perfect
controllability is that the final state of the controlled system, after the
actuation step, is statistically independent of the noise variable $Z$ given 
$X$ and $C$. In that case, the `information' $I(X^{\prime };Z|X,C)$ conveyed
in the form of noise from $Z$ to the controlled system is zero. Another
`common sense' interpretation of this result can be given if the quantity $
I(X^{\prime };Z|X,C)$ is instead viewed as representing the `information'
about $X^{\prime }$ that has been transferred to the non-controllable state $
Z$ in the form of `lost' correlations.

This analysis of control systems in terms of noise and information
protection is similar to that of error-correcting codes. The design of
error-correcting codes is closely related to that of control systems: the
information duplicated by a code, when corrupted by noise, is used to detect
errors (sensor step) which are then corrected by enacting specific
correcting or erasure actions (actuation step) \cite
{shannon1948,roman1992,cerf1997}. The analogy to error-correcting codes can
be strengthened even further if probabilities accounting for undetected and
uncorrected errors are modeled by means of communication channels similar to
the sensor and actuation channels. In this context, whether or not a
prescribed set of erasure actions is sufficient to correct for a particular
type of errors is determined by the control loss.

\subsection{\protect\medskip Observability}

The concept of observability is concerned with the issue of inferring the
state $X$ of the controlled system based on some knowledge or data of the
state provided by a measurement apparatus, taken here to correspond to $C$.
More precisely, a controlled system is termed \textit{perfectly observable}
if the sensor's transition matrix $p(c|x)$ maps no two values of $X$ to a
single observational output value $c$, or in other words if for all $c\in 
\mathcal{C}$ there exists only one value $x$ such that $p(x|c)=1$. As a
consequence, we have the following result \cite{lloyd1996}. (We omit the
proof which readily follows from well-known properties of entropy.)

\textit{Theorem 5.} A system with state variable $X$ is perfectly
observable, with respect to all observed value $c\in \text{supp}(C)$, if and
only if 
\begin{equation}
H(X|C)=\sum_{c\in \mathcal{C}}H(X|c)p_C(c)=0.  \label{pct1}
\end{equation}

The information-theoretic analog of a perfectly observable system is a 
\textit{lossless} communication channel $X\rightarrow Y$ characterized by $
H(X|Y)=0$ for all input distributions \cite{cover1991}. As a consequence of
this association, we interpret the conditional entropy $H(X|C)$ as the
information loss, or \textit{sensor loss}, of the sensor channel, denoted by 
$L_S$. We now extend our results on controllability into the domain of
observability. The first question that arises is, given the similarity
between the average control loss $L_C$ and the sensor loss, do we obtain
true results for observability by merely substituting $L_C$ by $L_S$ in
Theorems 2 and 3?

The answer is no: the fact that a communication channel is lossless has
nothing to do with the fact that it can be non-deterministic. An example of
such a channel is one that maps the singleton input set $\mathcal{X}=\{0\}$
to multiple instances of the output set $\mathcal{C}$ with equal
probabilities. This is clearly a non-deterministic channel, and yet since
there is only one possible value for $X$, the conditional entropy $H(X|c)$
must be equal to zero for all $c\in \mathcal{C}$. Hence, contrary to Theorem
2, the observation loss $L_S$ cannot be bounded above by the entropy of the
random variable responsible for the non-deterministic properties of the
sensor channel. However, we are not far from a similar result: by analyzing
the meaning of the sensor loss a bit further, the generalization of Theorem
2 for observability can in fact be derived using the `backward' version of
the sensor channel. More precisely, $L_S\leq H(Z_B)$ where $Z_B$ is now the
random variable associated with the purification of the transition matrix $
p(x|c)$. To prove this result, the reader may revise the proof of Theorem 2,
and replace the forward purification condition $H(C|X,Z)=0$ for the sensor
channel by its backward analog $H(X|C,Z_B)=0$.

To close this section, we present next what is left to generalization of the
results on controllability. One example aimed at illustrating the interplay
between the controllability and observability properties of a system is also
given.

\textit{Theorem 6.} If the state $X$ is perfectly observable, then $
I(X;Z|C)=0$. (The random variable $Z$ stands for the purification variable
of the `forward' sensor channel $p(c|x)$.)

\textit{Proof.} The proof is rather straightforward. Since $H(X|C)\geq
H(X|C,Z)$, the condition $L_S=0$ implies $H(X|C,Z)=0$. Thus by the chain
rule 
\begin{equation}
I(X;Z|C)=H(X|C)-H(X|C,Z),
\end{equation}
we conclude with $I(X;Z|C)=0$.\hfill$\blacksquare $

\textit{Corollary 7.} If $L_S=0$, then $I(X;C,Z)=I(X;C)$.

The interpretations of the two above results follow closely those given for
controllability. We will not discuss these results further except to mention
that, contrary to the case of controllability, $I(X;Z|C)=0$ is not a
sufficient condition for a system to be observable. This follows simply from
the fact that $I(X;Z|C)=0$ implies $H(X|C)=H(X|C,Z)$, and at this point the
purification condition $H(C|X,Z)=0$ for the sensor channel is of no help to
obtain $H(X|C)=0$.

\textit{Example 2.} Consider again the control system of Figure 3. Given the
actuation rules described by the \textsc{cnot} logical gate, it can be
verified easily that for $X=0$ or $1$, $H(X^{\prime }|x,C)=0$ and $
p(x^{\prime }|x)\neq 0$ for all $x^{\prime }$. Therefore, the controlled
system is completely and perfectly controllable. This implies, in
particular, that $\Delta H_{\text{open}}^c=\Delta H_{\text{closed}}^c=0$,
and that the final state of the controlled system may be actuated to a
single value with probability 1, as noted before. For the latter
observation, note that $X^{\prime }=x^{\prime }$ with probability 1 so long
as the initial state $X$ is known with probability 1 (perfectly observable).
In general, if a system is perfectly controllable (actuation property) 
\textit{and} perfectly observable (sensor property), then it is possible to
perfectly control its state to any desired value with vanishing probability
of error. In such a case, we can say that the system is \textit{closed-loop
controllable}.\hfill $\blacksquare $

\subsection{The case of continuous random variables}

The concept of a deterministic continuous random variable is somewhat
ill-defined, and, in any case, cannot be associated with the condition $
H(X)=0$ formally. (Consider, e.g., the peaked distribution $p(x)=\delta
(x-x_0)$ which is such that $H(X)=-\infty $.) To circumvent this difficulty,
controllability and observability for continuous random variables may be
extended via a quantization or coarse-graining of the relevant state spaces 
\cite{cover1991}. For example, a continuous-state system can be defined to
be perfectly controllable at $x$ if for every final destination $x^{\prime }$
there exists at least one control value $c$ which forces the system to reach
a small neighborhood of radius $\Delta >0$ around $x^{\prime }$ with
probability 1. Equivalently, $x$ can be termed perfectly controllable to
accuracy $\Delta $ if the variable $x^\Delta $ obtained by quantizing $
\mathcal{X}$ at a scale $\Delta $ is perfectly controllable. Similar
definitions involving quantized random variables can also be given for
observability. The recourse to the quantized description of continuous
variables has the virtue that $H(X^\Delta )$ and $H(X^\Delta |C^\Delta )$
are well-defined functions which cannot be infinite. It is also the natural
representation used for representing continuous-state models on computers.

\section{Stability and entropy reduction}

The emphasis in the previous section was on proving upper limits for the
control and the observation loss, and on finding conditions for which these
losses vanish. In this section, we depart from these quantities to focus our
attention on other measures which are interesting in view of the stability
properties of a controlled system. How can a system be stabilized to a
target state or a target subset (attractor) of states? Also, how much
information does a controller need to gather in order to achieve
successfully a stabilization procedure? To answer these questions, we first
propose an entropic criterion of stability, and justify its usefulness for
problems of control. In a second step, we investigate the quantitative
relationship between the closed-loop mutual information $I(X;C)$ and the
gain in stability which results from using information in a control process.

\subsection{Stochastic stability}

Intuitively, a stable system is a system which, when activated in the
proximity of a desired operating point, stays relatively close to that point
indefinitely in time, even in the presence of small perturbations. In the
field of control engineering, there exist several formalizations of this
intuition, some less stringent than others, whose range of applications
depend on theoretical as well as practical considerations. It would be
impossible, and, perhaps inappropriate, to review here all the definitions
of stability currently used in the study and design of control systems; for
our purposes, it suffices to say that a necessary condition for stabilizing
a dynamical system is to be able to decrease its entropy, or immunize it
from sources of entropy like those associated with environment noise, motion
instabilities, and incomplete specification of control conditions. This
entropic aspect of stabilization is implicit in almost all criteria of
stability insofar as a probabilistic description of systems focusing on sets
of responses, rather than on individual response one at a time, is adopted 
\cite{stengel1994,schlo1980,shaw1981,farmer1982}. In this sense, what is
usually sought in controlling a system is to confine its possible states,
trajectories or responses within a set as small as possible (low entropy
final state)\ starting from a wide range of initial states or initial
conditions (high entropy initial random state).

The fundamental role of entropy reduction in control suggests the two
following problems. First, given the initial state $X$ and its entropy $H(X)$
, a set of actuation subdynamics, and the type of controller (open- or
closed-loop), what is the maximum entropy reduction achievable during the
controlled transition from $X$ to $X^{\prime }$? Second, what is the
quantitative relationship between the maximal open-loop entropy reduction
and the closed-loop entropy reduction? Note that for control purposes it
does not suffice to reduce the entropy of $X^{\prime }$ conditionally on the
state of another system (the controller in particular). For instance, the
fact that $H(X^{\prime }|C)$ vanishes for a given controller acting on a
system does not imply by itself that $H(X^{\prime })$ must vanish as well,
or that $X^{\prime }$ is stabilized. What is required for control is that
actuators modify the dynamics of the system intended to be controlled by
acting directly on it, so as to reduce the marginal entropy $H(X^{\prime })$
. This unconditional aspect of stability has been discussed in more detail
in \cite{hugo12000,hugo22000}.

\subsection{Open-loop control optimality}

Using the concavity property of entropy, and the fact that $\Delta H_{\text{
open}}$ is upper bounded by the maximum of $\Delta H_{\text{open}}^c$ over
all control values $c$, we show in this section that the maximum decrease of
entropy achieved by a particular subdynamics of control variable 
\begin{equation}
\hat{c}=\arg \max_{c\in \mathcal{C}}\Delta H_{\text{open}}^c  \label{argm1}
\end{equation}
is \textit{open-loop optimal} in the sense that no random (i.e.,
non-deterministic) choice of the controller's state can improve upon that
decrease. More precisely, we have the following results. (Theorem 9 was
originally stated without a proof in \cite{hugo12000}.)

\textit{Lemma 8.} For any initial state $X$, the open-loop entropy reduction 
$\Delta H_{\text{open}}$ satisfies 
\begin{equation}
\Delta H_{\text{open}}\leq \Delta H_{\text{open}}^C,
\end{equation}
where 
\begin{eqnarray}
\Delta H_{\text{open}}^C &=&\sum_{c\in \mathcal{C}}p_C(c)\Delta H_{\text{
open }}^c  \nonumber \\
&=&H(X)-H(X^{\prime }|C)_{\text{open}}
\end{eqnarray}
with $\Delta H_{\text{open}}^c$ defined as in Eq.(\ref{opc}). The equality
is achieved if and only if $I(X^{\prime };C)=0$.

\textit{Proof.} Using the inequality $H(X^{\prime })\geq H(X^{\prime }|C)$,
we write directly 
\begin{eqnarray}
\Delta H_{\text{open}} &=&H(X)-H(X^{\prime })_{\text{open}}  \nonumber \\
&\leq &H(X)-H(X^{\prime }|C)_{\text{open}}.
\end{eqnarray}
Now, let us prove the equality part. If $C$ is statistically independent of $
X^{\prime }$, then $H(X^{\prime }|C)=H(X^{\prime })$, and 
\begin{equation}
\Delta H_{\text{open}}=\Delta H_{\text{open}}^C.
\end{equation}
Conversely, the above equality implies $H(X^{\prime }|C)=H(X^{\prime })$,
and thus we must have that $C$ is independent of $X^{\prime }$.\hfill $
\blacksquare $

\textit{Theorem 9.} The entropy reduction achieved by a set of actuation
subdynamics used in open-loop control is always such that 
\begin{equation}
\Delta H_{\text{open}}\leq \max_{c\in \mathcal{C}}\Delta H_{\text{open}}^c,
\label{opop}
\end{equation}
for all $p_X(x)$. The equality can always be achieved for the deterministic
controller $C=\hat{c}$, with $\hat{c}$ defined as in Eq.(\ref{argm1}).

\textit{Proof.} The average conditional entropy $H(X^{\prime }|C)$ is always
such that 
\begin{equation}
\min_{c\in \mathcal{C}}H(X^{\prime }|c)\leq \sum_{c\in \mathcal{C}
}p_C(c)H(X^{\prime }|c).
\end{equation}
Therefore, making use of the previous lemma, we obtain 
\begin{eqnarray}
\Delta H_{\text{open}} &\leq &\Delta H_{\text{open}}^C  \nonumber \\
&\leq &H(X)-\min_{c\in \mathcal{C}}H(X^{\prime }|c)  \nonumber \\
&=&\max_{c\in \mathcal{C}}\Delta H_{\text{open}}^c.
\end{eqnarray}
Also, note that if $C=\hat{c}$ with probability 1, then the two above
inequalities are saturated since in this case $I(X^{\prime };C)=0$ and $
\Delta H_{\text{open}}^C=\Delta H_{\text{open}}^{\hat{c}}$.\hfill$
\blacksquare $

An open-loop controller or a control strategy is called \textit{pure} if the
control random variable $C$ is deterministic, i.e., if it assumes only one
value with probability 1. An open-loop controller that is not pure is called 
\textit{mixed}. (We also say that a mixed controller activates a mixture of
control actions.) In view of these definitions, what we have just proved is
that a pure controller with $C=\hat{c}$ is necessarily optimal; any mixture
of the control variable either achieves the maximum entropy decrease
prescribed by Eq.(\ref{opop}) or yields a smaller value. As shown in the
next example, this is so even if the actuation subdynamics used in the
control process are deterministic.

\textit{Example 3.} For the \textsc{cnot} controller of Example 1, we noted
that $H(X^{\prime })_{\text{open}}=H(X)$, or equivalently that $\Delta H_{
\text{open}}=0$, only \textit{at best}. To be more precise, $\Delta H_{\text{
open}}=0$ only if a pure controller is used or if $H(X)=1$ bit (already at
maximum entropy). If the control is mixed, and if $H(X)<1$ bit, then $\Delta
H_{\text{open}}$ must necessarily be negative. This is so because
uncertainty as to which actuation rule is used must imply uncertainty as to
which state the controlled system is actuated to.\hfill $\blacksquare $

Note that purity alone is not a sufficient condition for open-loop
optimality, nor it is a necessary one in fact. To see this, note on the one
hand that a pure controller having 
\begin{equation}
C=\arg \min_{c\in \mathcal{C}}\Delta H_{\text{open}}^c
\end{equation}
with probability one is surely not optimal, unless all entropy reductions $
\Delta H_{\text{open}}^c$ have the same value. On the other hand, to prove
that a mixed controller can be optimal, note that if any subset $\mathcal{C}
_o\subseteq \mathcal{C}$ of actuation subdynamics is such that $p(x^{\prime
}|c)=p_{X^{\prime }}(x^{\prime })$, and $\Delta H_{\text{open}}^c$ assumes a
constant value for all $c\in \mathcal{C}_o$, then one can build an optimal
controller by choosing a non-deterministic distribution $p(c)$ with $\text{
supp}(C)=\mathcal{C}_o$.

\subsection{Closed-loop control optimality}

The distinguishing characteristic of an open-loop controller is that it
usually fails to operate efficiently when faced with uncertainty and noise.
An open-loop controller acting independently of the state of the controlled
system, or solely based on the statistical information provided by the
distribution $p_X(x)$, cannot reliably determine which control subdynamics
is to be applied in order for the initial (\textit{a priori} unknown) state $
X$ to be propagated to a given target state. Furthermore, an open-loop
control system cannot compensate actively in time for any disturbances that
add to the actuator's driving state (actuation noise). To overcome these
difficulties, the controller must be adaptive: it must be capable of
estimating the unpredictable features of the controlled system during the
control process, and must be able to use the information provided by
estimation to decide of specific control actions, just as in closed-loop
control.

A basic closed-loop controller was presented in Example 1. For this example,
we noted that the perfect knowledge of the initial state's value ($X=0$ or $
1 $) enabled the controller to decide which actuation subdynamics (identity
or permutation) is to be used in order to actuate the system to $X^{\prime
}=0$ with probability 1. The fact that the sensor gathers $I(X;C)=H(X)$ bits
of information during estimation is a necessary condition for this specific
controller to achieve $H(X^{\prime })_{\text{closed}}=0$, since having $
I(X;C)<H(X)$ may result in generating the value $X^{\prime }=1$ with
non-vanishing probability. In general, just as a subdynamics mapping the
input states $\{0,1\}$ to the single value $\{0\}$ would require no
information to force $X^{\prime }$ to assume the value $0$, we expect that
the closed-loop entropy reduction should not only depend on $I(X;C)$, the
effective information available to the controller, but should also depend on
the reduction of entropy attainable by open-loop control. The next theorem,
which constitutes the main result of this work, embodies exactly this
statement by showing that one bit of information gathered by the controller
has a maximum value of one bit in the improvement of entropy reduction that
closed-loop gives over open-loop control.

\textit{Theorem 10.} The amount of entropy 
\begin{equation}
\Delta H_{\text{closed}}=H(X)-H(X^{\prime })_{\text{closed}}
\end{equation}
that can be extracted from a system with given initial state $X$ by using a
closed-loop controller with fixed set of actuation subdynamics satisfies 
\begin{equation}
\Delta H_{\text{closed}}\leq \Delta H_{\text{open}}^{\max }+I(X;C).
\label{clot1}
\end{equation}
where 
\begin{equation}
\Delta H_{\text{open}}^{\max }=\max_{p_X(x)\in \mathcal{P},c\in \mathcal{C}
}\Delta H_{\text{open}}^c  \label{max1}
\end{equation}
is the maximum entropy decrease that can be obtained by (pure) open-loop
control over \textit{any} input distribution chosen in the set $\mathcal{P}$
of all probability distributions.

A proof of the result, based on the conservation of entropy for closed
systems, was given in \cite{hugo12000} following results found in \cite
{lloyd1989,caves1996}. Here, we present an alternative proof based on
conditional analysis which has the advantage over our previous work to give
some indications about the conditions for equality in (\ref{clot1}). Some of
these conditions are derived in the next section.

\textit{Proof.} Given that $\Delta H_{\text{open}}^{\max }$ is the optimal
entropy reduction for open-loop control over any input distribution, we can
write 
\begin{equation}
H(X^{\prime })_{\text{open}}\geq H(X)-\Delta H_{\text{open}}^{\max }.
\end{equation}
Now, using the fact that a closed-loop controller is formally equivalent to
an ensemble of open-loop controllers acting on the conditional supports $
\text{supp}(X|c)$ instead of $\text{supp}(X)$, we also have for all $c\in 
\mathcal{C}$ 
\begin{equation}
H(X^{\prime }|c)_{\text{closed}}\geq H(X|c)-\Delta H_{\text{open}}^{\max },
\end{equation}
and, on average, 
\begin{equation}
H(X^{\prime }|C)_{\text{closed}}\geq H(X|C)-\Delta H_{\text{open}}^{\max }.
\label{bd1}
\end{equation}
That $\Delta H_{\text{open}}^{\max }$ must enter in the lower bounds of $
H(X^{\prime })_{\text{open}}$ and $H(X^{\prime })_{\text{closed}}$ can be
explained in other words by saying that each conditional distribution $
p(x|c) $ is a legitimate input distribution for the initial state of the
controlled system. It is, in any cases, an element of $\mathcal{P}$. This
being said, notice now that $H(X^{\prime })\geq H(X^{\prime }|C)$ implies 
\begin{equation}
H(X^{\prime })_{\text{closed}}\geq H(X|C)-\Delta H_{\text{open}}^{\max }.
\end{equation}
Hence, we obtain 
\begin{eqnarray}
\Delta H_{\text{closed}} &\leq &H(X)-H(X|C)+\Delta H_{\text{closed}}^{\max }
\nonumber \\
&=&I(X;C)+\Delta H_{\text{closed}}^{\max },
\end{eqnarray}
which is the desired upper bound. To close the proof, note that $\Delta H_{
\text{open}}^{\max }$ cannot be evaluated using the initial distribution $
p_X(x)$ alone because the maximum reduction of entropy in open-loop control
starting from $p_X(x)$ may differ from the reduction of entropy obtained
when some actuation channel is applied in closed-loop to $p(x|c)$. See \cite
{hugo22000} for a specific example of this.\hfill $\blacksquare $

The above theorem enables us to finally understand all the results of
Example 1. As noted already, since the actuation subdynamics consist of
permutations, we have $\Delta H_{\text{open}}^{\max }=0$ for any
distribution $p_X(x)$. Thus, we should have $\Delta H_{\text{closed}}\leq
I(X;C)$. For the particular case studied where $C=X$, the controller is
found to be \textit{optimal}, i.e., it achieves the maximum possible entropy
reduction $\Delta H_{\text{closed}}=I(X;C)$. This proves, incidentally, that
the bound of inequality (\ref{clot1}) is tight. In general, we may define a
control system to be optimal in terms of information if the gain in
stability obtained by substracting $H(X^{\prime })_{\text{open}}$ from $
H(X^{\prime })_{\text{closed}}$ is exactly equal to the sensor mutual
information $I(X;C)$. Equivalently, a closed-loop control system is optimal
if its \textit{efficiency }$\eta $, defined by 
\begin{equation}
\eta =\frac{H(X^{\prime })_{\text{open}}-H(X^{\prime })_{\text{closed}}}{
I(X;C)},
\end{equation}
is equal to 1.

Having determined that optimal controllers do exist, we now turn to the
problem of finding general conditions under which a given controller is
found to be either optimal ($\eta =1$) or sub-optimal ($\eta <1$). By
analyzing thoroughly the proof of Theorem 10, one finds that the assessment
of the condition $I(X^{\prime };C)=0$, which was not a sufficient condition
for open-loop optimality, is again not sufficient here to conclude that a
closed-loop controller is optimal. This comes as a result of the fact that
not all control subdynamics applied in a closed-loop fashion are such that $
\Delta H_{\text{closed}}^c=\Delta H_{\text{open}}^{\max }$ in general.
Therefore the average final condition entropy $H(X^{\prime }|C)_{\text{
closed }}$ need not necessarily be equal to the bound imposed by inequality (
\ref{bd1}). However, in a scenario where the entropy reductions $\Delta H_{
\text{ open}}^c$ and $\Delta H_{\text{closed}}^c$ are both equal to a
constant for all control subdynamics, then we effectively recover an analog
of the open-loop optimality condition, namely that a zero mutual information
between the controller and the controlled system \textit{after} actuation is
a necessary and sufficient condition for optimality.

\textit{Theorem 11.} Under the condition that, for all $c\in \mathcal{C}$, 
\begin{equation}
\Delta H_{\text{open}}^c=\Delta H_{\text{closed}}^c=\Delta H,  \label{const1}
\end{equation}
where $\Delta H$ is a constant, then a closed-loop controller is optimal if
and only if $I(X^{\prime };C)=0$.

\textit{Proof.} To prove the sufficiency part of the theorem, note that the
constancy condition (\ref{const1}) implies that the minimum for $H(X^{\prime
})_{\text{open}}$ equals $H(X)-\Delta H$. Similarly, closed-loop control
must be such that 
\begin{equation}
H(X^{\prime }|C)_{\text{closed}}=H(X|C)-\Delta H.  \label{const2}
\end{equation}
Combining these results with the fact that $I(X^{\prime };C)=0$, or
equivalently that 
\begin{equation}
H(X^{\prime })_{\text{closed}}=H(X^{\prime }|C)_{\text{closed}},
\end{equation}
we obtain 
\begin{eqnarray}
H(X^{\prime })_{\text{open}}^{\min }-H(X^{\prime })_{\text{closed}}
&=&H(X)-H(X|C)  \nonumber \\
&=&I(X;C).  \label{opt1}
\end{eqnarray}
To prove the converse, namely that optimality under condition (\ref{const1})
implies $I(X^{\prime };C)=0$, notice that Eq.(\ref{const2}) leads to 
\begin{eqnarray}
H(X^{\prime })_{\text{open}}^{\min }-H(X^{\prime }|C)_{\text{closed}}
&=&H(X)-H(X|C)  \nonumber \\
&=&I(X;C).
\end{eqnarray}
Hence, given that we have optimality, i.e., given Eq.(\ref{opt1}), then $
X^{\prime }$ must effectively be independent of $C$.\hfill$\blacksquare $

\textit{Example 4.} Consider again the now familiar \textsc{cnot}
controller. Let us assume that instead of the perfect sensor channel $C=X$,
we have a binary symmetric channel such that $p(c=x|x)=1-e$ and $p(c=x\oplus
1|x)=e$ where $0\leq e\leq 1$, i.e., an error in the transmission occurs
with probability $e$ \cite{cover1991}. The mutual information for this
channel is readily calculated to be 
\begin{eqnarray}
I(X;C) &=&H(C)-\sum_{x\in \{0,1\}}p(x)H(C|x)  \nonumber \\
&=&H(C)-H(e),
\end{eqnarray}
where 
\begin{equation}
H(e)=-e\log e-(1-e)\log (1-e)
\end{equation}
is the binary entropy function. By proceeding similarly as in Example 1, the
distribution of the final controlled state can be calculated. The solution
is $p_{X^{\prime }}(0)=1-e$ and $p_{X^{\prime }}(1)=e$, so that $H(X^{\prime
})=H(e)$ and 
\begin{equation}
\Delta H_{\text{closed}}=H(X)-H(e).
\end{equation}
By comparing the value of $\Delta H_{\text{closed}}$ with the mutual
information $I(X;C)$ (recall that $\Delta H_{\text{open}}^{\max }=0$), we
arrive at the conclusion that the controller is optimal for $e=0$, $e=1$
(perfect sensor channel), and for $H(X)=1$ (maximum entropy state). In going
through more calculations, it can be shown that these cases of optimality
are all such that $I(X^{\prime };C)=0$.\hfill $\blacksquare $

\subsection{Continuous-time limit}

To derive a differential analog of the closed-loop optimality theorem for
systems evolving continuously in time, one could try to proceed as follows:
sample the state, say $X(t)$, of a controlled system at two time instants
separated by some (infinitesimal) interval $\Delta t$, and from there
directly apply inequality (\ref{clot1}) to the open- and closed-loop entropy
reductions associated with the two end-points $X(t)$ and $X(t+\Delta t)$
using $I(X(t);C(t))$ as the information gathered at time $t$. However sound
this approach might appear, it unfortunately proves to be inconsistent for
many reasons. First, although one may obtain well-defined rates for $H(X(t))$
in the open- or closed-loop regime, the quantity 
\begin{equation}
\lim_{\Delta t\rightarrow 0}\frac{I(X(t);C(t))}{\Delta t}  \label{pr1}
\end{equation}
does not constitute a rate, for $I(X(t);C(t))$ is not a differential element
which vanishes as $\Delta t$ approaches $0$. Second, our very definition of
open-loop control, namely the requirement that $I(X;C)$ be equal to $0$
prior to actuation, fails to apply for continuous-time dynamics. Indeed,
open-loop controllers operating continuously in time must always be such
that $I(X(t);C(t))\neq 0$ if purposeful control is to take place. Finally,
are we allowed to extend a result derived in the context of a Markovian or
memoryless model of controllers to sampled continuous-time processes, even
if the sampled version of such processes has a memoryless structure? Surely,
the answer is no.

To overcome these problems, we suggest the following conditional version of
the optimality theorem. Let $X(t-\Delta t)$, $X(t)$ and $X(t+\Delta t)$ be
three consecutive sampled points of a controlled trajectory $X(t)$. Also,
let $C(t-\Delta t)$ and $C(t)$ be the states of the controller during the
time interval in which the state of the controlled system is estimated. (The
actuation step is assumed to take place between the time instants $t$ and $
t+\Delta t$.) Then, by redefining the entropy reductions as conditional
entropy reductions following 
\begin{equation}
\Delta H^t=H(X(t)|C^{t-\Delta t})-H(X(t+\Delta t)|C^{t-\Delta t}),
\end{equation}
where $C^t$ represents the control history up to time $t$, we must have 
\begin{equation}
\Delta H_{\text{closed}}^t\leq \Delta H_{\text{open}}^t+I(X(t);C(t)|C^{t-
\Delta t}).
\end{equation}
Note that by thus conditioning all quantities with $C^{t-\Delta t}$, we
extend the applicability of the closed-loop optimality theorem to any class
of control processes, memoryless or not. Now, since 
\begin{equation}
I(X(t-\Delta t);C(t-\Delta t)|C^{t-\Delta t})=0
\end{equation}
by the definition of the mutual information, we also have 
\begin{eqnarray}
\Delta H_{\text{closed}}^t &\leq &\Delta H_{\text{open}}^t+I(X(t);C(t)|C^{t-
\Delta t})  \nonumber \\
&&-I(X(t-\Delta t);C(t-\Delta t)|C^{t-\Delta t}).
\end{eqnarray}
As a result, by dividing both sides of the inequality by $\Delta t$, and by
taking the limit $\Delta t\rightarrow 0$, we obtain the rate equation 
\begin{equation}
\dot{H}_{\text{closed}}\leq \dot{H}_{\text{open}}+\dot{I}
\end{equation}
if, indeed, the limit exists. This equation relates the rate at which the
conditional entropy $H(X(t)|C^{t-\Delta t})$ is dissipated in time with the
rate at which the conditional mutual information $I(X(t);C(t)|C^{t-\Delta
t}) $ is gathered upon estimation. The difference between the above
information rate and the previous pseudo-rate reported in Eq.(\ref{pr1})
lies in the fact that $I(X(t);C(t)|C^{t-\Delta t})$ represents the
differential information gathered during the \textit{latest} estimation
stage of the control process. It does not include past correlations induced
by the control history $C^{t-\Delta t}$. This sort of conditioning allows,
in passing, a perfectly meaningful re-definition of open-loop control in
continuous-time, namely $\dot{I}=0$, since the only correlations between $
X(t)$ and $C(t)$ which can be accounted for in the absence of direct
estimation are those due to the past control history.

\section{Applications}

\subsection{Proportional controllers}

There are several controllers in the real world which have the character of
applying a control signal with amplitude proportional to the distance or
error between some estimate $\hat{X}$ of the state $X$, and a desired target
point $x^{*}$. In the control engineering literature, such controllers are
designated simply by the term \textit{proportional }controllers \cite
{stengel1994}. As a simple version of a controller of this type, we study in
this section the following system: 
\begin{eqnarray}
X^{\prime } &=&X-C  \nonumber \\
C &=&\hat{X}-x^{*},  \label{claw}
\end{eqnarray}
with all random variables assuming values on the real line. For simplicity,
we set $x^{*}=0$ and consider two different estimation or sensor channels
defined mathematically by 
\begin{equation}
C_\Delta =\hat{X}=\left( \left\lfloor \frac X\Delta \right\rfloor +\frac
12\right) \Delta ,  \label{cgch}
\end{equation}
and 
\begin{equation}
C_Z=\hat{X}=X+Z,  \label{gsch}
\end{equation}
where $Z\sim \mathcal{N}(0,N)$ (Gaussian distribution with zero mean and
variance $N$). The first kind of estimation, Eq.(\ref{cgch}), is a
coarse-grained measurement of $X$ with a grid of size $\Delta $; it
basically allows the controller to `see' $X$ within a precision $\Delta $,
and selects the middle coordinate of each cell of the grid as the control
value for $C_\Delta $. The other sensor channel represented by the control
state $C_Z$ is simply the Gaussian channel with noise variance $N$.

Let us start our study of the proportional controller by considering the
coarse-grained sensor channel first. If we assume that $X\sim \mathcal{U}
(0,\varepsilon )$ (uniform distribution over an interval $\varepsilon $
centered around $0$), and pose that $\varepsilon /\Delta $ is an integer,
then we must have 
\begin{equation}
I(X;C_\Delta )=\log (\varepsilon /\Delta ).  \label{coinf}
\end{equation}
Now, to obtain $p_{X^{\prime }}(x^{\prime })_{\text{closed}}$, note that the
conditional random variables $X|c$ defined by conditional analysis are all
uniformly distributed over non-overlapping intervals of width $\varepsilon
/\Delta $, and that, moreover, all of these intervals must be moved under
the control law around $X^{\prime }=0$ without deformation. Hence, $
X^{\prime }\sim \mathcal{U}(0,\Delta )$, and 
\begin{eqnarray}
\Delta H_{\text{closed}} &=&\log \varepsilon -\log \Delta  \nonumber \\
&=&\log (\varepsilon /\Delta ).
\end{eqnarray}
These results, combined with the fact that $\Delta H_{\text{open}}^{\max }=0$
, prove that the coarse-grained controller is always optimal, at least
provided again that $\varepsilon $ is a multiple of $\Delta $.

In the case of the Gaussian channel, the situation for optimality is
different. Under the application of the estimation law (\ref{gsch}), the
final state of the controlled system is 
\begin{equation}
X^{\prime }=X-C=X-(X+Z)=-Z,
\end{equation}
so that $X^{\prime }\sim Z$. This means that if we start with $X\sim 
\mathcal{N}(0,P)$, then 
\begin{eqnarray}
\Delta H_{\text{closed}} &=&\frac 12\log (2\pi eP)-\frac 12\log (2\pi eN) 
\nonumber \\
&=&\frac 12\log \frac PN,
\end{eqnarray}
and 
\begin{equation}
I(X;C_Z)=\frac 12\log \left( 1+\frac PN\right) .
\end{equation}
Again, $\Delta H_{\text{open}}^{\max }=0$ (recall that $\Delta H_{\text{open}
}^{\max }$ does not depend on the choice of the sensor channel), and so we
conclude that optimality is achieved only in the limit where the
signal-to-noise ratio goes to infinity. Non-optimality, for this control
setup, can be traced back to the presence of some overlap between the
different conditional distributions $p(x|c)$ which is responsible for the
mixing upon application of the control. As $P/N\rightarrow \infty $, the
`area' covered by the overlapping regions decreases, and so is $I(X^{\prime
};C)$. Based on this observation, we have attempted to change the control
law slightly so as to minimize the mixing in the control while keeping the
overlap constant and found that complete optimality for the Gaussian channel
controller can be achieved if the control law is modified to 
\begin{equation}
X^{\prime }=X-\gamma C,
\end{equation}
with a \textit{gain} parameter $\gamma $ set to 
\begin{equation}
\gamma =\frac P{P+N}.
\end{equation}
This controller can readily be verified to be optimal.

\subsection{Noisy control of chaotic maps}

The second application is aimed at illustrating the closed-loop optimality
theorem in the context of a controller restricted to use entropy-increasing
actuation dynamics, as is often the case in the control of chaotic systems.
To this end, we consider the feedback control scheme proposed by Ott,
Grebogi and Yorke (OGY) \cite{ott1990} as applied to the \textit{logistic
map } 
\begin{equation}
x_{n+1}=f(r_n,x_n)=r_nx_n(1-x_n),  \label{logmap}
\end{equation}
where $x_n\in [0,1]$, and $r_n\in [0,4]$, $n=0,1,2,\ldots $. In a nutshell,
the OGY control method consists in setting the control parameter $r_n$ at
each time step $n$ according to 
\begin{eqnarray}
r_n &=&r+\delta r_n  \nonumber \\
\delta r_n &=&-\gamma (c_n-x^{*})  \label{ogy}
\end{eqnarray}
whenever the estimated state $c_n=\hat{x}_n$ falls into a small control
region $D$ in the vicinity of a target point $x^{*}$. This target state is
usually taken to be an unstable fixed point satisfying the equation $
f(r,x^{*})=x^{*}$, where $f(r,x^{*})$ is the unperturbed map having $r_n=r$
as a constant control parameter. Moreover, the gain $\gamma $ is fixed so as
to ensure that the trajectory $\{x_n\}_{n=0}^\infty $ is stable under the
control action. (See \cite{shin1993,schust1995} for a derivation of the
stability conditions for $\gamma $ based on linear analysis, and \cite
{shin1995,boc2000} for a review of the field of chaotic control.)

\begin{figure}[t]

\epsfig{file=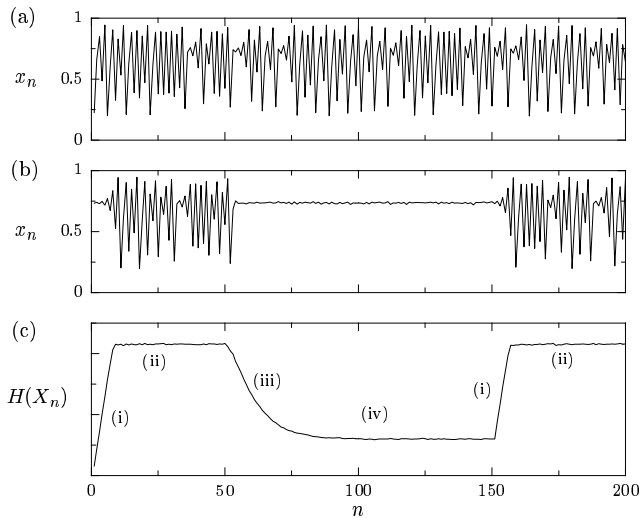,width=3.35in,clip=}

\caption{(a) Typical uncontrolled trajectory of the
logistic map with $r=3.7825$. (b)  Controlled trajectory which results from applying the 
OGY feedback control at time $n=50$. Note the instant resurgence of instability
as the control is switched off at $n=150$. 
The gain for this simulation was set to  $\gamma=-7.0$, and $D=[0.725,0.745]$.  
(c) Entropy $H(X_n)$ (in arbitrary units) associated with the position of the controlled
system versus time (see text).}

\end{figure}

Figure 4 illustrates the effect of OGY controller when applied to the
logistic map. The plot of Figure 4a shows a typical chaotic trajectory
obtained by iterating the dynamical equation (\ref{logmap}) with $
r_n=r=3.7825$. Note on this plot the presence of non-recurring oscillations
around the unstable fixed point $x^{*}(r)=(r-1)/r\simeq 0.7355$. Figure 4b
shows the orbit of the same initial point $x_0$ now stabilized by the OGY
controller around $x^{*}$ for $n\in [50,150]$. For this latter simulation,
and more generally for any initial points in the unit interval, the
controller is able to stabilize the state of the logistic map in some region
surrounding $x^{*}$, provided that $\gamma $ is a stable gain, and that the
sensor channel is not too noisy. To evidence the stability properties of the
controller, we have calculated the entropy $H(X_n)$ by constructing a
normalized histogram $p_{X_n}(x_n)$ of the positions of a large ensemble of
trajectories ($\sim 10^4$) starting at different initial points. The result
of this numerical computation is shown in Figure 4c. On this graph, one can
clearly distinguish four different regimes in the evolution of $H(X_n)$,
numbered from (i) to (iv), which mark four different regimes of dynamics:

(i) \textit{Chaotic motion with constant }$r$: Exponential divergence of
nearby trajectories initially located in a very small region of the state
space. The slope of the linear growth of entropy, the signature of chaos 
\cite{beck1993,vito1999}, is probed by the value of the Lyapunov exponent 
\begin{equation}
\lambda (r)=\lim_{N\rightarrow \infty }\frac 1N\sum_{n=0}^{N-1}\ln \left|
\left. \frac{\partial f(r,x)}{\partial x}\right| _{x_n}\right| .
\label{lyap1}
\end{equation}

(ii) \textit{Saturation}: At this point, the distribution of positions $
p_{X_n}(x_n)$ for the chaotic system has reached a limiting or equilibrium
distribution which nearly fills all the unit interval.

(iii) \textit{Transient stabilization}:~When the controller is activated,
the set of trajectories used in the calculation of $H(X_n)$ is compressed
around $x^{*}$ exponentially rapidly in time.

(iv) \textit{Controlled regime}: An equilibrium situation is reached whereby 
$H(X_n)$ stays nearly constant. In this regime, the system has been
controlled down to a given residual entropy which specifies the size of the
basin of control, i.e., the average distance from $x^{*}$ to which $x_n$ has
been controlled.

\begin{figure*}[t]

\epsfig{file=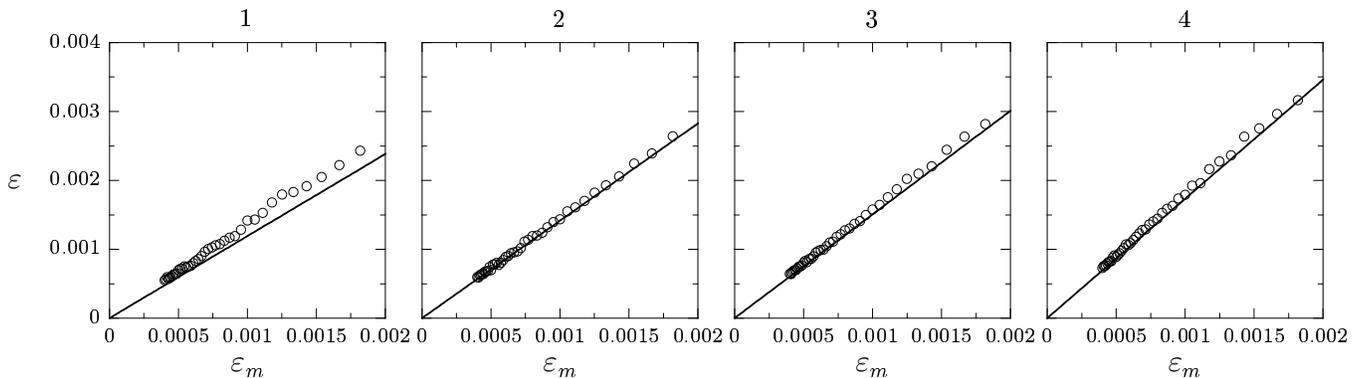,width=\textwidth,clip=}

\caption{(Data points) Control interval $\varepsilon$ as a function of the effective
coarse-grained interval of measurement $\varepsilon_m$ for four
different target points. (Solid line) Optimal linear relationship predicted by the
closed-loop optimality theorem. The values of $r$ and the Lyapunov exponents
$\lambda^*$ associated with the target points are listed in Table 1 and
displayed in Figure 6.}

\end{figure*}

It is the size of the basin of control, and, more precisely, its dependence
on the amount of information provided by the sensor channel which is of
interest to us here. In order to study this dependence, we have simulated
the OGY controller, and have compared the value of the residual entropy $
H(X_n)$ for two types of sensor channel:\ the coarse-grained channel $
C_n=C_\Delta (X_n)$, and the Gaussian channel $C_n=C_Z(X_n)$.

In the case of the coarse-grained channel, we have found that the
distribution of $X_n$ in the controlled regime was well approximated by a
uniform distribution of width $\varepsilon $ centered around the target
point $x^{*}$. Thus, the indicator value for the size of the basin of
control is taken to correspond to 
\begin{equation}
\varepsilon =e^{H(X_n)},  \label{contint}
\end{equation}
which, according to the closed-loop optimality theorem, must be such that 
\begin{equation}
\varepsilon \geq e^{\lambda ^{*}}\varepsilon _m,  \label{cgopt1}
\end{equation}
where $\lambda ^{*}$ is the Lyapunov exponent associated with the $r$ value
of the unperturbed logistic map, and where $\varepsilon _m$ is the
coarse-grained measurement interval or precision of the sensor channel. (All
logarithms are in natural base in this section.) To understand the above
inequality, note that a uniform distribution for $X_n$ covering an interval
of size $\delta $ must stretch by a factor $e^{\lambda (r)}$ after one
iteration of the map with parameter $r$. This follows from the fact that $
\lambda (r)$ corresponds to an entropy rate of the dynamical system \cite
{beck1993,vito1999} (see also \cite{shaw1981,farmer1982}), and holds in an
average sense inasmuch as the support of $X_n$ is not too small or does not
cover the entire unit interval. Now, for open-loop control, it can be seen
that if $\lambda (r)>0$ for all admissible control values $r$, then no
control of the state $X_n$ is possible, and the optimal control strategy
must consist in using the smallest Lyapunov exponent $\lambda _{\min }$
available in order to achieve 
\begin{eqnarray}
\Delta H_{\text{open}}^{\max } &=&H(X_n)-H(X_{n+1})_{\text{open}}  \nonumber
\\
&=&\ln \delta -\ln e^{\lambda _{\min }}\delta  \nonumber \\
&=&-\lambda _{\min }<0.
\end{eqnarray}
In the course of the simulations, we noticed that only a very narrow range
of $r$ values were actually used in the controlled regime, which means that $
\Delta H_{\text{open}}^{\max }$ can be taken for all purposes to be equal to 
$-\lambda ^{*}$. At this point, then, we need only to use expression (\ref
{coinf}) for the mutual information of the coarse-grained channel,
substituting $\Delta $ with $\varepsilon _m$, to obtain 
\begin{equation}
\Delta H_{\text{closed}}\leq -\lambda ^{*}+\ln (\varepsilon /\varepsilon _m).
\end{equation}
This expression yields the aforementioned inequality by posing $\Delta H_{
\text{closed}}=0$ (controlled regime).

\begin{figure}[t]

\epsfig{file=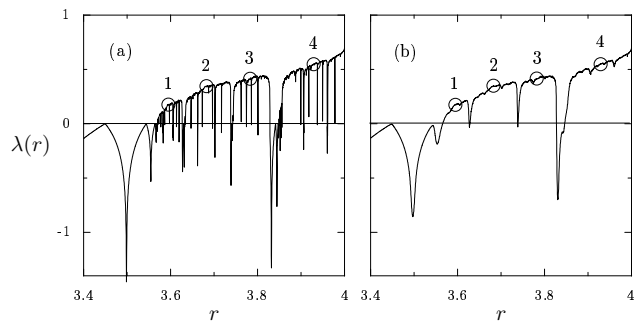,width=3.3in,clip=}

\caption{(a) Lyapunov spectrum $(r,\lambda(r))$ of the logistic map. 
The positive Lyapunov exponents associated with the four target points listed
in Table 1 are located by the circles. The set of $r$ values used 
during the control spans approximately the diameter of the circles.
Note that the few negative values of $\lambda(r)$ close to the $\lambda^*$'s are 
effectively suppressed by the noise in the sensor channel. This is evidenced
by the graph of (b) which was obtained by computing the sum (\ref{lyap1})
up to $N=2\times 10^4$ with an additive noise component of very small amplitude.
See \protect{\cite{crutch1982,crutch21982}} for more details on this point.}

\end{figure}

\begin{table}[t] \centering
\caption{Characteristics of the four target points.}
\begin{tabular}{c|ccc}
\hline\hline
Target point & $x^{*}$ & $r$ & $\lambda ^{*}$ (base $e$) \\ \hline
$1$ & $0.7218$ & $3.5950$ & $0.1745$ \\ 
$2$ & $0.7284$ & $3.6825$ & $0.3461$ \\ 
$3$ & $0.7356$ & $3.7825$ & $0.4088$ \\ 
$4$ & $0.7455$ & $3.9290$ & $0.5488$ \\ \hline\hline
\end{tabular}
\end{table}

\begin{figure*}[t]

\epsfig{file=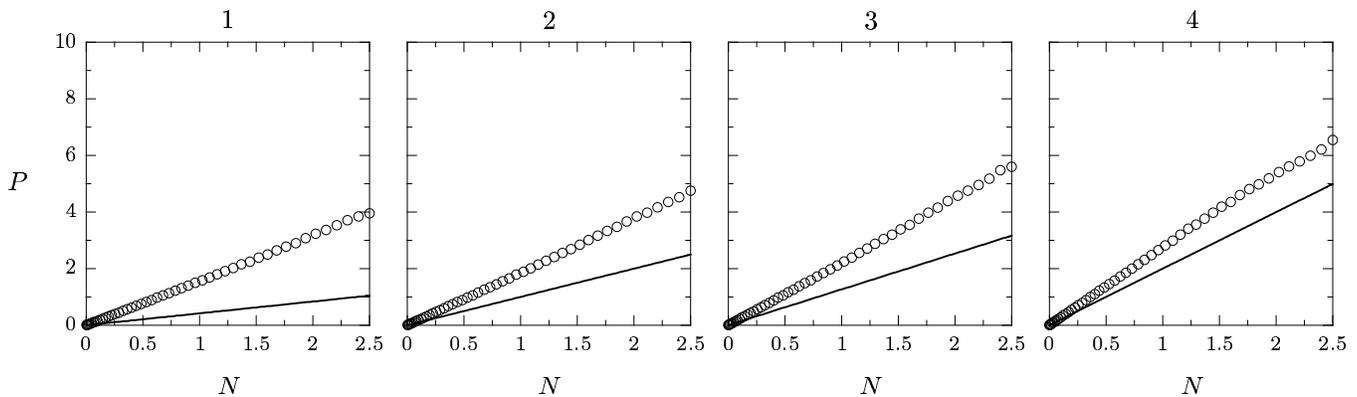,width=\textwidth,clip=}

\caption{(Data points) Dispersion $P$ characterizing the basin of attraction of the controlled
system as a function of the noise power $N$ introduced in the Gaussian
sensor channel. The horizontal and vertical axes are to be rescaled by a factor
$10^{-5}$. (Solid line) Optimal lower bound.}

\end{figure*}

The plots of Figure 5 present our numerical calculations of $\varepsilon $
as a function of $\varepsilon _m$. Each of these plots has been obtained by
calculating Eq.(\ref{contint}) using the entropy of the normalized histogram
of the positions of about $10^4$ different controlled trajectories. Other
details about the simulations may be found in the caption. What
differentiates the four plots is the fixed point to which the ensemble of
trajectories have been stabilized, and, accordingly, the value of the
Lyapunov exponent $\lambda ^{*}$ associated to $x^{*}(r)$. These are listed
in Table 1 and illustrated in Figure 6. One can verify on the plots of
Figure 5 that the points of $\varepsilon $ versus $\varepsilon _m$ all lie
above the critical line (solid line in the graphs) which corresponds to the
optimality prediction of inequality (\ref{cgopt1}). Also, the relatively
small departure of the numerical data from the optimal prediction shows that
the OGY controller with the coarse-grained channel is nearly optimal with
respect to the entropy criterion. This may be explained by noticing that
this sort of controller complies with all the requirements of the first
class of linear proportional controllers studied previously. Hence, we
expect it to be optimal for all precision $\varepsilon _m$, although the
fact must be considered that $\Delta H_{\text{open}}^{\max }=-\lambda ^{*}$
is only an approximation. In reality, not all points are controlled with the
same parameter $r$ for a given value of $\varepsilon _m$, as shown in Figure
6. Moreover, how $\varepsilon $ is calculated explicitly relies on the
assumption that the distribution for $X_n$ is uniform. This assumption has
been verified numerically; yet, it must also be regarded as an
approximation. Taken together, these two approximations may explain the
observed deviations of $\varepsilon $ from its optimal value.

For the Gaussian channel, optimality is also closely related to our results
about proportional controllers. The results of our simulations, for this
type of channel, indicated that the normalized histogram of the controlled
positions for $X_n$ is very close to a normal distribution with mean $x^{*}$
and variance $P$. As a consequence, we now consider the variance $P$, which
for Gaussian random variables is given by 
\begin{equation}
P=\frac{e^{2H(X_n)}}{2\pi e},
\end{equation}
as the correlate of the size of the basin of control. For this quantity, the
closed-loop optimality theorem with $\Delta H_{\text{closed}}=0$ yields 
\begin{equation}
P\geq (e^{2\lambda ^{*}}-1)N,  \label{optp}
\end{equation}
where $N$ is the variance of the zero-mean Gaussian noise perturbing the
sensor channel.

In Figure 7, we have displayed our numerical data for $P$ as a function of
the noise power $N$. The solid line gives the optimal relationship which
results from taking equality in the above expression, and from substituting
the Lyapunov exponent associated with one of the four stabilized points
listed in Table 1. From the plots of this figure, we verify again that $P$
is lower bounded by the optimal value predicted analytically. However, now
it can be seen that $P$ deviates significantly from its optimal value,
making clear that the OGY controller driven by the Gaussian noisy sensor
channel is not optimal (except in the trivial limit where $N\rightarrow 0$).
This is in agreement with our proof that linear proportional controllers
with Gaussian sensor channel are not optimal in general. On the plots of
Fig.~7, it is quite remarkable to see that the data points all converge to
straight lines. This suggests that the mixing induced by the controller, the
source of non-optimality, can be accounted for simply by modifying our
inequality for $P$ so as to obtain 
\begin{equation}
P=(e^{2\lambda ^{\prime }}-1)N.
\end{equation}
The new exponent $\lambda ^{\prime }$ can be interpreted as an \textit{\
effective} Lyapunov exponent; its value is necessarily greater than $\lambda
^{*}$, since the chaoticity properties of the controlled system are enhanced
by the mixing effect of the controller.

\section{Concluding remarks}

\subsection{Control and thermodynamics}

The reader familiar with thermodynamics may have noted a strong similarity
between the functioning of a controller, when viewed as a device aimed at
reducing the entropy of a system, and the thought experiment of Maxwell
known as the Maxwell's demon paradox \cite{leff1990}. Such a similarity was
already noted in the Introduction section of this work. In the case of
Maxwell's demon, the system to be controlled or `cooled' is a volume of gas;
the entropy to be reduced is the equilibrium thermodynamic entropy of the
gas; and the `pieces' of information gathered by the controller (the demon)
are the velocities of the atoms or molecules constituting the gas. When
applied to this scheme, our result on closed-loop optimality can be
translated into an absolute limit to the ability of the demon, or any
control devices, to convert heat to work. Indeed, consider a feedback
controller operating in a cyclic fashion on a system in contact with a heat
reservoir at temperature $T$. According to Clausius law of thermodynamic 
\cite{reif1965}, the amount of heat $\Delta Q_{\text{closed}}$ extracted by
the controller upon reducing the entropy of the controlled system by a
concomitant amount $\Delta H_{\text{closed}}$ must be such that 
\begin{equation}
\Delta Q_{\text{closed}}=(k_BT\ln 2)\Delta H_{\text{closed}}.
\end{equation}
In the above equation, $k_B$ is the Boltzmann constant which provides the
necessary conversion between units of energy (Joule) and units of
temperature (Kelvin); the constant $\ln 2$ arises because physicists usually
prefer to express logarithms in base $e$. From the closed-loop optimality
theorem, we then write 
\begin{eqnarray}
\Delta Q_{\text{closed}} &\leq &(k_BT\ln 2)[\Delta H_{\text{open}}^{\max
}+I(X;C)]  \nonumber \\
&=&\Delta Q_{\text{open}}^{\max }+(k_BT\ln 2)I(X;C),
\end{eqnarray}
where $\Delta Q_{\text{open}}^{\max }=(k_BT\ln 2)\Delta H_{\text{open}
}^{\max }$. This limit should be compared with analogous results found by
other authors on the subject of thermodynamic demons (see, e.g., the
articles reprinted in \cite{leff1990}, and especially Szilard's analysis of
Maxwell's demon \cite{szilard1929} which contains many premonitory insights
about the use of information in control.)

It should be remarked that the connection between the problem of Maxwell's
demon, thermodynamics, and control is effective only to the extent that
Clausius law provides a link between entropy and the physically measurable
quantity that is energy. But, of course, the notion of entropy is a more
general notion than what is implied by Clausius law; it can be defined in
relation to several situations which have no direct relationship whatsoever
with physics (e.g., coding theory, rate distortion theory, decision theory).
This versatility of entropy is implicit here. Our results do not rely on
thermodynamic principles, or even physical principles for that matter, to be
true. They constitute valid results derived in the context of a general
model of control processes whose precise nature is yet to be specified.

\subsection{Entropy and optimal control theory}

Consideration of entropy as a measure of dispersion and uncertainty led us
to choose this quantity as a control function of interest, but other
information-theoretic quantities may well have been chosen instead if
different control applications require so. From the point of view of optimal
control theory, all that is required is to minimize a desired performance
criterion (a cost or a Lyapunov function), such as the distance to a target
point or the energy consumption, while achieving some desired dynamic
performance (stability) using a set of permissible controls \cite
{stengel1994,saridis1995}. For example, one may be interested in maximizing $
\Delta H_{\text{closed}}$ instead of minimizing this quantity if
destabilization (anti-control) or mixing is an issue \cite{alessand1999}. As
other examples, let us mention the minimization of the relative entropy
distance between the distribution of the state of a controlled system and
some target distribution \cite{beghi1996}, the problem of coding \cite
{ahl1998}, as well as the minimization of rate-like functions in decision or
game theory \cite
{jianhua1988,kelly1956,bellman1957,weidemann21969,guiasu1970,middleton1996,cover1991}
.

\subsection{Future work}

Many questions pertaining to issues of information and control remain at
present unanswered. We have considered in this paper the first level of
investigation of a much broader and definitive program of research aimed at
providing information-theoretic tools for the study of general control
systems, such as those involving many interacting components, as well as
controllers exploiting non-Markovian features of dynamics (e.g., memory,
learning, and adaptation). In a sense, what we have studied can be compared
with the memoryless channel of information theory; what is needed in the
future is something like a control analog of network information theory.
Work is ongoing along this direction.

\begin{acknowledgments}

H.T. would like to thank P. Dumais for correcting a preliminary version of
the manuscript, S. Patagonia for inspiring thoughts, an anonymous referee
for useful suggestions, and especially V. Poulin for her always critical
comments. Many thanks are also due to A.-M. Tremblay for the permission to
access the supercomputing facilities of the CERPEMA at the Universit\'{e} de
Sherbrooke. S.L. was supported by DARPA, by ARDA/ARO and by the NSF under
the National Nanotechnology Initiative. H.T. was supported by NSERC
(Canada), and by a grant from the d'Arbeloff Laboratory for Information
Systems and Technology at MIT during the initial phase of this work.

\end{acknowledgments}


\begin{thebibliography}{99}
\bibitem{note1}  From the viewpoint of information, the simplification $S=A=C
$ amounts to a situation whereby the sensor is connected to the actuator by
a noiseless communication channel described by a one-to-one mapping between
the input $S$ and the output $A$ of the controller.
\bibitem{note3}  How the distribution for $C$ is chosen depends, in general,
on which value $X=x$ is characterized as being controllable; hence the fact
that we must consider a conditional distribution $p(c|x)$ for $C$ and not
the unconditional distribution $p_C(c)$.
\end{thebibliography}
\end{document}